\documentclass[runningheads]{llncs}

\usepackage[T1]{fontenc}
\usepackage{graphicx}
\usepackage{array}
\usepackage{arydshln}
\usepackage[inline]{enumitem}
\usepackage{adjustbox}
\newlength\myHeight
\newlength\myWidth
\usepackage{booktabs}
\usepackage{adjustbox}
\usepackage{multirow}
\usepackage{svg}
\usepackage{xstring}
\usepackage{float}
\usepackage{adjustbox}
\usepackage{orcidlink}
\definecolor{mforestgreen}{HTML}{228B22}

\definecolor{myblue}{HTML}{0036c7}
\definecolor{mygreen}{HTML}{228B22}
\definecolor{myorange}{HTML}{FF5630}

\usepackage{ifthen}
\usepackage{ulem}
\usepackage{cleveref}
\newboolean{mocommentenabled}
\newboolean{onlysuggestedtext}
\newcommand{\enablemocomment}{\setboolean{mocommentenabled}{true}}

\newcommand{\mocomment}[2]{%
  \ifthenelse{\boolean{mocommentenabled}}{%
    \ifthenelse{\boolean{onlysuggestedtext}}{%
      \ifthenelse{\equal{#2}{}}{#1}{#2}%
    }{%
      \ifthenelse{\equal{#2}{}}{#1}{\sout{#1}\textcolor{blue}{#2}}%
    }%
  }{#1}%
}
\enablemocomment
\begin{document}

\title{Cross-Dataset Bloom Question Classification: Supervised Models and Prompted LLMs}
\titlerunning{Cross-Dataset Bloom Question Classification}

\author{
  Abdolali Faraji\inst{1}\thanks{These authors contributed equally to this work.} \orcidlink{0000-0002-3557-9345} \and
  Mohammadreza Molavi\inst{1}\textsuperscript{*} \orcidlink{0009-0006-0423-0729} \and
  Zohreh Rasoulkhani \inst{2} \orcidlink{0009-0009-0628-1695} \and
  Mohammadreza Tavakoli \inst{1} \orcidlink{0000-0002-7368-0794} \and
  Gábor Kismihók \inst{1} \orcidlink{0000-0003-3758-5455}
}

\authorrunning{A. Faraji et al.}

\institute{Leibniz Information Centre for Science and Technology (TIB)\\
    \email{\{abdolali.faraji, mohammadreza.molavi, reza.tavakoli, gabor.kismihok\}@tib.eu}
\and
University of Genoa\\
    \email{zrasoulkhani@gmail.com}}

\maketitle

\begin{abstract}
Automatic Bloom’s taxonomy classification of assessment questions can substantially reduce instructor workload, but labeling is subjective and teacher-dependent. Prior machine learning (ML) and deep learning (DL) approaches reported strong within-dataset results, yet were rarely evaluated in cross-dataset settings, leaving real-world generalizability unclear; meanwhile, LLM effectiveness for Bloom question classification has not been systematically studied. We evaluated the cross-dataset generalization of existing ML/DL methods and assessed LLMs with multiple prompting strategies on five datasets; the best prompting strategy combined in-context examples with course-specific action verbs. Supervised ML/DL models degraded substantially on unseen datasets, whereas LLMs were more stable, suggesting a robust alternative across diverse educational contexts. Based on the best prompting strategy, we also presented a lightweight UI that supports instructors in automatically classifying large question banks; a usability study indicated low workload and high usability.

\end{abstract}

\section{Introduction}
Bloom’s taxonomy~\cite{bloom1956taxonomy} and its revised formulation~\cite{anderson2001taxonomy} have long served as foundational frameworks for organizing learning objectives and assessment items according to levels of cognitive complexity. By structuring cognitive processes from lower-order to higher-order skills, Bloom’s taxonomy supports the systematic design, analysis, and alignment of instructional activities and assessments.

Despite its importance, manually classifying assessment questions based on Bloom’s taxonomy is time-consuming, especially for large item banks or repeated course offerings~\cite{yahya2012bloom}. This has motivated interest in automating Bloom-level classification to reduce instructor workload and improve efficiency. A key challenge, however, is that Bloom labeling is inherently context- and teacher-specific: the same question may be interpreted differently depending on instructional goals, course content, or instructor perspective~\cite{scaria2024automated}. Developing automated tools that perform reliably across teachers and contexts thus remains a non-trivial problem.

Most existing efforts to automate Bloom-level question classification rely on ML and DL methods~\cite{gani2023bloom,mohammed2020question}. While these approaches often achieve high performance, evaluations are generally conducted using train-test splits from the same dataset. Consequently—given the context- and instructor-specific nature of Bloom labeling discussed above—the generalizability of these models to other datasets or instructors remains unclear, limiting their practical utility.

More recently, large language models (LLMs) have emerged as an alternative to traditional ML and DL approaches, demonstrating strong performance across a wide range of text classification tasks~\cite{almatrafi2025leveraging}. However, despite their growing use in educational applications, their potential for Bloom–based classification of assessment questions has not yet been examined in a focused manner.

\begin{figure}[t]
  \centering
  \includegraphics[width=0.93\linewidth]{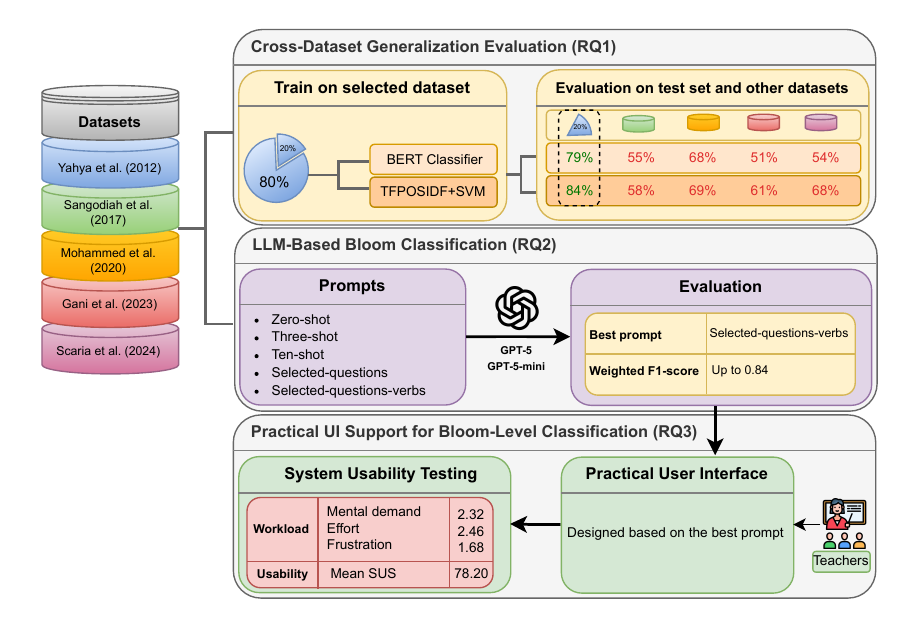}
  \caption{Paper overview. The section corresponding to RQ1 illustrates cross-dataset evaluation using one dataset as the training source, highlighting performance drops when testing on other datasets. The RQ2 section presents the evaluation of LLMs across datasets under different prompting strategies and the selection of the best-performing prompt. The RQ3 section shows the usability study of the UI built on this prompt.}
  \label{fig:toc}
\end{figure}

Motivated by these gaps, this study investigated automatic Bloom classification of assessment questions. First, we examined the cross-dataset generalization of existing ML and DL approaches, assessing how well models trained in one context transferred to unseen datasets. Second, we evaluated the performance of LLMs for Bloom-based question classification across datasets and prompting strategies. Finally, to support teachers in automating the classification of large question banks, we designed and evaluated the usability of a user interface (UI) based on the best-performing method. Accordingly, we addressed the following research questions, which are also depicted in \Cref{fig:toc}:

RQ1: How well do developed ML and DL models for Bloom-based question classification generalize across unseen datasets?

RQ2: How do LLMs perform on Bloom classification of assessment questions?

RQ3: How can Bloom-based question classification be supported through a practical UI?

\section{Related Work}

This section positions our work by summarizing prior research. We group prior work into two methodological categories: ML/DL methods and more recent studies that leverage LLMs for cognitive classification.

\subsection{ML and DL Approaches for Bloom Classification}

Early Bloom question-classification work used traditional ML. Yahya et al.~\cite{yahya2012bloom} framed the task as supervised learning with SVMs and analyzed the impact of frequency features and stopword removal. Mohammed and Omar~\cite{mohammed2020question} proposed POS-aware TF--IDF that emphasizes verbs, combined it with word embeddings, and evaluated several classifiers (SVM, logistic regression, KNN), improving over standard TF--IDF baselines. Other work explored alternative approaches: Das et al.~\cite{das2020identification} applied LLDA for \textit{multi-class} Bloom classification, while Wang et al.~\cite{wang2021towards} proposed a \textit{weakly supervised} method to reduce labeling needs.

More recent work has increasingly adopted DL models. Shaikh et al.~\cite{shaikh2021bloom} applied LSTM networks with pretrained word embeddings to classify both assessment questions and course learning outcomes. Gani et al.~\cite{gani2023bloom} investigated convolutional neural networks combined with various pretrained embeddings, including both non-contextual (Word2Vec, GloVe, FastText) and contextual representations (BERT, RoBERTa, ELECTRA), reporting their strongest results when using CNNs with RoBERTa embeddings. Das et al.~\cite{das2020identification} further evaluated transformer-based models by fine-tuning BERT for multi-class Bloom classification, demonstrating substantial gains over traditional methods.

Despite strong reported performance, evaluations typically used random within-dataset train--test splits. As a result, models trained on one dataset were rarely tested on different datasets, leaving generalization across instructors and educational contexts largely unexplored and limiting applicability to unseen data.

\subsection{Large Language Models for Cognitive Classification}

Recent work has explored LLMs for cognitive classification in educational settings. Scaria et al.~\cite{scaria2024automated} focused on Bloom-level question generation and used Bloom classification only for evaluating generated questions, offering an initial (but not dedicated) look at the classification task.

Almatrafi et al.~\cite{almatrafi2025leveraging} studied GPT-4 for Bloom-based categorization of course learning outcomes, comparing multiple prompting strategies (zero-shot, few-shot, and chain-of-thought). Faraby et al.~\cite{faraby2024analysis} analyzed ChatGPT for educational question classification using the Graesser taxonomy, which characterizes question types by underlying cognitive processes rather than Bloom’s levels.

Overall, prior LLM work focuses on learning outcomes, uses other taxonomies, or uses Bloom classification only to evaluate generated questions. Systematic multi-dataset evaluations of LLMs for Bloom question classification remain limited, suggesting the area is still in its infancy.

\section{Method}
This section provides an overview of our methodology, including datasets, models, evaluation procedures, and results. The code, prompts, and datasets are available in our repository\footnote{\url{https://gitlab.com/zrasoulkhani/bloom-classification-aied2026}}.
\subsection{Datasets}

This study uses five Bloom’s taxonomy–annotated question datasets, totaling 4,179 questions. Four datasets consist of human-authored assessment questions that have been frequently adopted in Bloom classification research: Yahya et al. (2012) \cite{yahya2012bloom}, Sangodiah et al. (2017) \cite{sangodiah2017taxonomy}, Mohammed and Omar (2020) \cite{mohammed2020question}, and Gani et al. (2023) \cite{gani2023bloom}, containing 600, 415, 126, and 1,200 questions, respectively. In addition to human-generated questions, we incorporate an LLM-generated question dataset from Scaria et al. (2024) \cite{scaria2024automated} (1,838 questions). This dataset is included to reflect the growing prevalence of AI-generated educational content in contemporary learning environments.

While all datasets provide Bloom-based labels, some follow the original Bloom taxonomy, whereas others use the revised Bloom taxonomy. To ensure consistency and comparability across datasets, we map all labels to the revised Bloom cognitive levels; the resulting class distribution is shown in \Cref{tab:class-dist}. 

\begin{table}[ht]
\centering
\caption{Distribution of questions across revised Bloom levels in the five datasets.}
\label{tab:class-dist}
\resizebox{0.95\textwidth}{!}{%
\begin{tabular}{lrrrrrrr}  % r instead of c for right-aligned numbers
\hline
Dataset & Remember & Understand & Apply & Analyze & Evaluate & Create & Total \\
\hline
Yahya et al. (2012) & 100 & 100 & 100 & 100 & 100 & 100 & 600 \\
Sangodiah et al. (2017)  & 50  & 135 & 72  & 56  & 45  & 57  & 415 \\
Mohammed and Omar (2020) & 22  & 20  & 15  & 19  & 29  & 21  & 126 \\
Gani et al. (2023) & 669 & 107 & 100 & 149 & 99  & 76  & 1\,200 \\
Scaria et al. (2024) (LLM-generated) & 218 & 412 & 338 & 220 & 344 & 306 & 1\,838 \\
\hline
Total          & 1\,059 & 774 & 625 & 544 & 617 & 560 & 4\,179 \\
\hline
\end{tabular}%
}
\end{table}

\subsection{Cross-Dataset Generalization Evaluation (RQ1)}
\subsubsection{Classification Models}
To examine the generalizability of Bloom’s taxonomy question classification models, we evaluate two representative approaches from prior work, covering both traditional ML and DL paradigms.

\paragraph{TFPOSIDF+SVM.}
This model is an SVM-based classifier that represents questions using TF--IDF features weighted by part-of-speech (POS) information. In this approach, words are assigned different weights according to their POS tags, with higher importance given to verbs, as they provide strong indicators of Bloom’s cognitive levels~\cite{mohammed2020question}.

\paragraph{Fine-tuned BERT.}
As a deep learning reference model, we employ a BERT-based text classification model fine-tuned on Bloom-labeled data in our experiments.  We adopt the model architecture and hyperparameter configuration reported in prior work that applied BERT to Bloom-based classification~\cite{almatrafi2025leveraging}.

\subsubsection{Evaluation Protocol}
For each dataset, we perform a cross-dataset generalization evaluation. Specifically, one dataset is selected as the source dataset and randomly split into 80\% training and 20\% test sets. Models are trained on the training portion, and performance is first measured on the test set using weighted F1-score, reflecting the evaluation protocol commonly reported in prior studies.
The trained models are then evaluated on the remaining four datasets, which are treated as unseen test sets. This procedure is repeated for each dataset, such that every dataset serves once as the training source and multiple times as an unseen test set.

\subsubsection{Results and Discussion.}
\Cref{tab:generalization-results} reports the weighted F1-scores for TFPOSIDF+SVM and fine-tuned BERT across the five datasets. Diagonal values correspond to performance on the test split of the training dataset, while the other values represent cross-dataset test results. For TFPOSIDF+SVM, test scores range from 0.68 to 0.82 depending on the training dataset, but performance decreases on unseen datasets by an average of 0.25, ranging between 0.40 and 0.70. Similarly, fine-tuned BERT achieves test scores between 0.76 and 0.91, which drop on average by 0.28 when evaluated on other datasets, ranging from 0.31 to 0.76.

These results highlight the inherent subjective and context-dependent nature of Bloom's classification of questions: models trained on questions from one instructor or context do not generalize reliably to unseen datasets from different contexts. The only exception observed in our results occurs when TFPOSIDF+SVM is trained on the Sangodiah dataset and tested on the Gani dataset, showing a small increase of 0.02 in weighted F1-score. For BERT, no increase is observed, but the performance drop is only 0.04, which is considerably smaller than the average cross-dataset decrease of 0.28. This asymmetry suggests that the diversity and context present in the Gani questions are already represented in the Sangodiah dataset, allowing better transfer in this direction.

\begin{table}[ht]
\centering
\caption{Cross-dataset generalization results (weighted F1-score).}
\label{tab:generalization-results}
\resizebox{0.95\textwidth}{!}{%
\begin{tabular}{llrrrrrr}  % r instead of c for right-aligned numbers
\hline
\raisebox{-5pt}{Model} & \raisebox{-5pt}{Train dataset} & \multicolumn{5}{c}{Test datasets} \\
\cline{3-7}
  &  & Yahya & Sangodiah & Mohammed & Gani & Scaria \\
\hline
     & Yahya     & \textbf{0.79} & 0.55  & 0.68  & 0.51 & 0.54 \\
     & Sangodiah  & 0.43 & \textbf{0.68} & 0.54  & 0.70 & 0.52  \\
TFPOSIDF+SVM  & Mohammed   & 0.46 & 0.44 & \textbf{0.73}  & 0.55 & 0.53 \\
     & Gani    & 0.40 & 0.59 & 0.49  & \textbf{0.82} & 0.41   \\
     & Scaria   & 0.41 & 0.46  & 0.51  & 0.54 & \textbf{0.78}  \\
\hdashline
    & Yahya     & \textbf{0.84} & 0.58  & 0.69  & 0.61 & 0.68 \\
    & Sangodiah  & 0.47 & \textbf{0.80} & 0.48  & 0.76 & 0.53  \\
Fine-tuned BERT & Mohammed  & 0.46 & 0.37 & \textbf{0.76}  & 0.31 & 0.36  \\
    & Gani    & 0.56 & 0.68 & 0.59  & \textbf{0.91} & 0.52   \\
    & Scaria    & 0.55 & 0.53  & 0.58  & 0.60 & \textbf{0.84}  \\

\end{tabular}%
}
\end{table}

\subsection{LLM-Based Bloom Classification (RQ2)}

\subsubsection{Large Language Models}
To examine the potential of LLMs for Bloom’s taxonomy question classification, we evaluate two models: GPT-5 and GPT-5-mini. Both models are used in a prompt-based inference setting via the Batch API with default configurations and do not undergo task-specific fine-tuning. Instead, they rely on in-context instructions and examples to assign Bloom’s cognitive levels to input questions. GPT-5 represents a higher-capacity model, while GPT-5-mini serves as a smaller and more economical alternative, allowing us to examine performance across both state-of-the-art and cost-efficient LLM settings.

\subsubsection{Prompting Strategies}
To classify questions with LLMs, we evaluated five different prompting strategies. All prompts were designed to produce outputs in JSON format, enabling automatic evaluation across datasets.

\begin{itemize}
    \item \textbf{Zero-shot:} The model is provided only the question and asked to assign a Bloom cognitive level.
    \item \textbf{Three-shot:} Three randomly selected examples per cognitive level, drawn from the same dataset, are provided to give minimal contextual guidance.
    \item \textbf{Ten-shot:} Ten randomly selected examples per cognitive level, again drawn from the same dataset, are included to provide richer contextual information.
    \item \textbf{Selected-questions:} Up to ten examples per level are manually selected to better represent the diversity of question types. Selection was performed jointly by two experts with at least five years of teaching and assessment experience for each dataset.
    \item \textbf{Selected-questions-verbs:} Building on the \textit{selected-questions} prompt, we added the curated sets of key action verbs selected by the experts for each Bloom cognitive level within each dataset to the prompt.

\end{itemize}

\subsubsection{Evaluation Protocol}
For each dataset, LLMs classify all questions using each prompting strategy, and predicted Bloom levels are evaluated against gold labels with weighted F1-score.

\subsubsection{Results and Discussion}
\Cref{tab:llm-results} reports the weighted F1-scores of GPT-5 and GPT-5-mini across all datasets and prompting strategies. The results show that prompting strategies based on carefully selected examples are particularly effective. Specifically, the \textit{selected-questions-verbs} prompt achieves the strongest performance, reaching weighted F1-scores of up to 0.84 on the Yahya dataset. This indicates that selecting representative examples, together with incorporating level-specific action verbs, provides valuable guidance for Bloom classification.

More generally, the use of in-context examples consistently improves model performance. Both three-shot and ten-shot prompting outperform zero-shot pro\-mpting across datasets, suggesting that additional contextual information helps to align model predictions with Bloom’s cognitive levels. Nevertheless, zero-shot prompting still yields reasonable performance, with GPT-5 achieving weighted F1-scores around 0.75 on most datasets, indicating that LLMs retain useful classification capability even in the absence of examples. At the model level, across all prompting strategies and datasets, GPT-5 consistently performs better than GPT-5-mini; however, the performance gap remains modest, making GPT-5-mini a viable alternative when computational efficiency or cost is a concern, with an average weighted F1-score only 0.03 lower.

Compared to the supervised models evaluated in RQ1, LLM performance shows substantially less sensitivity to dataset shifts, indicating stronger cross-dataset robustness. Although LLMs do not surpass the within-dataset test performance of fine-tuned models, their comparatively stable performance across diverse datasets highlights their potential as a practical alternative for Bloom classification in heterogeneous educational settings. This robustness is particularly valuable given the subjective and context-dependent nature of Bloom’s taxonomy, where labeled data from a single instructional context may not generalize reliably to unseen sources.

\begin{table}[ht]
\centering
\caption{Weighted F1-score results of GPT-5 and GPT-5-mini on various datasets.}
\label{tab:llm-results}
\resizebox{0.95\textwidth}{!}{%
\begin{tabular}{llrrrrrr}  % r instead of c for right-aligned numbers
\hline
\raisebox{-5pt}{LLM} & \raisebox{-5pt}{Prompt} & \multicolumn{5}{c}{Test datasets} \\
\cline{3-7}
 &  & Yahya & Sangodiah & Mohammed & Gani & Scaria \\
\hline
   & selected-questions-verbs & \textbf{0.84} & \textbf{0.76} & \textbf{0.81} & \textbf{0.83} & \textbf{0.82} \\
   & selected-questions  & 0.79 & 0.75  & 0.80  & 0.80  & \textbf{0.82} \\
gpt-5 & 10-shot  & 0.79  & 0.74  & 0.79  & 0.75  & \textbf{0.82}  \\
   & 3-shot & 0.76 & 0.74 & 0.76 & 0.75 & 0.80  \\
   & 0-shot & 0.75 & 0.71 & 0.76 & 0.75 & 0.75  \\
\hdashline
   & selected-questions-verbs & 0.80 & 0.71 & 0.77 & 0.80 & 0.81  \\
   & selected-questions  & 0.79 & 0.71  & 0.77  & 0.77  & 0.80 \\
gpt-5-mini & 10-shot  & 0.75  & 0.69  & 0.76  & 0.73  & 0.79  \\
   & 3-shot & 0.74 & 0.67 & 0.73 & 0.72 & 0.79  \\
   & 0-shot & 0.74 & 0.65 & 0.72 & 0.71 & 0.73 \\

\end{tabular}%
}
\end{table}

\subsection{User Interface to Support Bloom Classification (RQ3)}
Bloom-level question classification is difficult to do manually at scale, as instructors often need to label large numbers of questions~\cite{yahya2012bloom}. To support this process, we developed a lightweight tool based on our best-performing prompting strategy (\textit{selected-questions-verbs}). Instructors can provide a small set of example questions per Bloom level for their specific context (e.g., course, topic, or learning objectives); the tool extracts key action verbs from these examples and uses them---together with the examples themselves---to construct the prompt for classifying new questions. Users can then upload question sets (CSV/Excel) and receive Bloom-level predictions in a structured output.

To assess usability, we conducted a user study with $N=50$ participants recruited via Prolific\footnote{\url{https://www.prolific.com}}. Participants (1) interacted with the UI for 10 minutes for one of their real courses and added Bloom-level examples, (2) answered three NASA-TLX items (mental demand, effort, frustration; 1--5 scale)~\cite{hart1988development}, and (3) completed the System Usability Scale (SUS)~\cite{brooke1996sus}. Participants reported low perceived workload (means: mental demand $2.32$, effort $2.46$, frustration $1.68$), and the tool achieved very good usability (mean SUS $=78.2$, SD $=14.07$)\footnote{The detailed study data are available online: \url{https://bit.ly/4asf6ZA}.}.

\section{Conclusion}
In this work, we examined the generalizability of prior Bloom-level question classification methods and the potential of LLMs for this task. Our cross-dataset experiments on five datasets show that models trained on a single dataset suffer substantial performance drops when applied to unseen datasets, highlighting limitations in their real-world applicability. In contrast, LLMs demonstrate more stable performance across datasets, with GPT-5 achieving weighted F1-scores up to 0.84 under the best prompt, indicating their potential as a more robust solution for Bloom-based question classification. To support practical use, we implemented a UI based on the \textit{selected-questions-verbs} workflow; a user study ($N=50$) reported low workload (NASA-TLX means: mental demand $2.32$, effort $2.46$, frustration $1.68$) and very good usability (mean SUS $=78.2$).

There are a few aspects that warrant further investigation. For cross-dataset evaluation, we only tested a subset of ML and DL models, and additional models could be explored. Regarding LLMs, more models, particularly open-source alternatives, could be evaluated to assess robustness and cost-effectiveness. Finally, while our UI design was tested with pilot users, feedback from real instructors would provide more precise insights for practical deployment

\bibliographystyle{splncs04}
\bibliography{References}

@article{shaikh2021bloom,
  title={Bloom’s learning outcomes’ automatic classification using LSTM and pretrained word embeddings},
  author={Shaikh, Sarang and Daudpotta, Sher Muhammad and Imran, Ali Shariq},
  journal={Ieee Access},
  volume={9},
  pages={117887--117909},
  year={2021},
  publisher={IEEE}
}

@article{das2020identification,
  title={Identification of cognitive learning complexity of assessment questions using multi-class text classification},
  author={Das, Syaamantak and Mandal, Shyamal Kumar Das and Basu, Anupam},
  journal={Contemporary Educational Technology},
  volume={12},
  number={2},
  pages={ep275},
  year={2020},
  publisher={Bastas}
}

@article{faraby2024analysis,
  title={Analysis of llms for educational question classification and generation},
  author={Al Faraby, Said and Romadhony, Ade and others},
  journal={Computers and Education: Artificial Intelligence},
  volume={7},
  pages={100298},
  year={2024},
  publisher={Elsevier}
}

@article{mohammed2020question,
  title={Question classification based on Bloom’s taxonomy cognitive domain using modified TF-IDF and word2vec},
  author={Mohammed, Manal and Omar, Nazlia},
  journal={PloS one},
  volume={15},
  number={3},
  pages={e0230442},
  year={2020},
  publisher={Public Library of Science San Francisco, CA USA}
}

@incollection{yahya2012bloom,
  title={Bloom’s taxonomy--based classification for item bank questions using support vector machines},
  author={Yahya, Anwar Ali and Toukal, Zakaria and Osman, Addin},
  booktitle={Modern advances in intelligent systems and tools},
  pages={135--140},
  year={2012},
  publisher={Springer}
}

@article{sangodiah2017taxonomy,
  title={Taxonomy Based Features in Question Classification Using Support Vector Machine.},
  author={Sangodiah, Anbuselvan and Ahmad, Rohiza and WAN AHMAD, WAN FATIMAH},
  journal={Journal of Theoretical \& Applied Information Technology},
  volume={95},
  number={12},
  year={2017}
}

@article{almatrafi2025leveraging,
  title={Leveraging generative AI for course learning outcome categorization using Bloom's taxonomy},
  author={Almatrafi, Omaima and Johri, Aditya},
  journal={Computers and Education: Artificial Intelligence},
  volume={8},
  pages={100404},
  year={2025},
  publisher={Elsevier}
}

@inproceedings{scaria2024automated,
  title={Automated educational question generation at different bloom’s skill levels using large language models: Strategies and evaluation},
  author={Scaria, Nicy and Dharani Chenna, Suma and Subramani, Deepak},
  booktitle={International Conference on Artificial Intelligence in Education},
  pages={165--179},
  year={2024},
  organization={Springer}
}

@article{gani2023bloom,
  title={Bloom’s Taxonomy-based exam question classification: The outcome of CNN and optimal pre-trained word embedding technique},
  author={Gani, Mohammed Osman and Ayyasamy, Ramesh Kumar and Sangodiah, Anbuselvan and Fui, Yong Tien},
  journal={Education and Information Technologies},
  volume={28},
  number={12},
  pages={15893--15914},
  year={2023},
  publisher={Springer}
}

@inproceedings{wang2021towards,
  title={Towards blooms taxonomy classification without labels},
  author={Wang, Zichao and Manning, Kyle and Mallick, Debshila Basu and Baraniuk, Richard G},
  booktitle={International conference on artificial intelligence in education},
  pages={433--445},
  year={2021},
  organization={Springer}
}

@article{bloom1956taxonomy,
  title={Taxonomy of educational objectives: the classification of educational goals},
  author={BS, BLOOM},
  journal={Handbook; Cognitive domain},
  volume={1},
  year={1956},
  publisher={McKay}
}

@book{anderson2001taxonomy,
  title={A taxonomy for learning, teaching, and assessing: A revision of Bloom's taxonomy of educational objectives: complete edition},
  author={Anderson, Lorin W and Krathwohl, David R},
  year={2001},
  publisher={Addison Wesley Longman, Inc.}
}

@inproceedings{brooke1996sus,
  title={SUS: A "quick and dirty" usability scale},
  author={Brooke, John},
  booktitle={Usability evaluation in industry},
  editor={Jordan, Patrick W and Thomas, Bruce and McClelland, Ian L and Weerdmeester, Bernard},
  pages={189--194},
  year={1996},
  publisher={Taylor \& Francis}
}

@incollection{hart1988development,
  title={Development of NASA-TLX (Task Load Index): Results of empirical and theoretical research},
  author={Hart, Sandra G and Staveland, Lowell E},
  booktitle={Advances in Psychology},
  volume={52},
  pages={139--183},
  year={1988},
  publisher={North-Holland}
}
\end{document}